# Effects of Pressure and Doping on Ruddlesden-Popper phases La$_{n+1}$Ni$_n$O$_{3n+1}$

Mingxin Zhang[1], Cuiying Pei[1], Qi Wang[1,2], Yi Zhao[1], Changhua Li[1], Weizheng Cao[1], Shihao Zhu[1], Juefei Wu[1], Yanpeng Qi[1,2,3*]

1. School of Physical Science and Technology, ShanghaiTech University, Shanghai 201210, China
2. ShanghaiTech Laboratory for Topological Physics, ShanghaiTech University, Shanghai 201210, China
3. Shanghai Key Laboratory of High-resolution Electron Microscopy, ShanghaiTech University, Shanghai 201210, China

* Correspondence should be addressed to Y.P.Q. (qiyp@shanghaitech.edu.cn)

**ABSTRACT**

**Recently the discovery of superconductivity with a critical temperature $T_c$ up to 80 K in Ruddlesden-Popper phases La$_{n+1}$Ni$_n$O$_{3n+1}$ ($n$ = 2) under pressure has garnered considerable attention. Up to now, the superconductivity was only observed in La$_3$Ni$_2$O$_7$ single crystal grown with the optical-image floating zone furnace under oxygen pressure. It remains to be understood the effect of chemical doping on superconducting La$_3$Ni$_2$O$_7$ as well as other Ruddlesden-Popper phases. Here, we systematically investigate the effect of external pressure and chemical doping on polycrystalline Ruddlesden-Popper phases. Our results demonstrate the application of pressure and doping effectively tunes the transport properties of Ruddlesden-Popper phases. We find pressure-induced superconductivity up to 86 K in La$_3$Ni$_2$O$_7$ polycrystalline sample, while no signatures of superconductivity are observed in La$_2$NiO$_4$ and La$_4$Ni$_3$O$_{10}$ systems under high pressure up to 50 GPa. Our study sheds light on the exploration of high-$T_c$ superconductivity in nickelates.**

## Introduction

The layered mixed valence nickelates have received enormous attention due to the similar crystal and electronic structures of cuprates over the past few decades[1-3]. The Ruddlesden-Popper (RP) phase $La_{n+1}Ni_nO_{3n+1}$ is one of the typical nickelates, which consist of $n$ layers of perovskite-type $LaNiO_3$, separated by single rocksalt-type $LaO$ layer along the crystallographic *c*-axis direction. Although extensive efforts have been devoted to searching for superconductivity in the RP phase akin to cuprates, experimental breakthroughs in this direction have not been achieved until recently. In 2019, *Li* et al. discovered superconductivity with $T_c \simeq 9–15$ K in the thin films of hole-doped infinite-layer nickelates $Nd_{1-x}Sr_xNiO_2$, which are obtained from the RP phase through removing two apical oxygens by a topochemical reduction method[4-6]. Afterward, several other alkaline-earth metals doped rare-earth nickelates such as $Ln_{1-x}Ae_xNiO_2$ (Ln = La, Pr; Ae = Sr, Ca)[7-10] and $Nd_6Ni_5O_{12}$ [11]have been found to exhibit superconductivity with $T_c$ up to 17 K. Interestingly, by applying pressure, the $T_c$ of $Pr_{0.82}Sr_{0.18}NiO_2$ could enhance monotonically to 31 K without showing the trend towards saturation[9].

Note that all the above superconductivity was achieved in thin films. Very recently, the superconductivity near 80 K has been found in bulk single crystals of bilayer RP phase $La_3Ni_2O_7$ under high pressure[12]. At ambient pressure, $La_3Ni_2O_7$ is a metal with a density-wave-like phase transition at about 120 K[13-15]. Pressure-induced superconductivity in $La_3Ni_2O_7$ accompanies a structural transition from *Amam* phase to *Fmmm* phase at around 14 GPa. After this work, several theoretical works were proposed to understand the mechanism of the high $T_c$ superconductivity in $La_3Ni_2O_7$ [16-28]. Besides, zero resistance was reached in electrical transport measurements with a hydrostaticity-improved diamond anvil cell (DAC)[29, 30].

Although the discovery of superconductivity in nickelates is exciting, there still are several open questions that have to be clarified: (1) Do polycrystalline samples show superconductivity under high pressure as single crystals? (2) Could the $T_c$ be improved after chemical doping in the $La_3Ni_2O_7$ sample? (3) Can we observe superconductivity in other members of RP phases (e.g., $La_2NiO_4$, $La_4Ni_3O_{10}$)? Motivated by the above issues, we synthesized a series of polycrystalline $La_{n+1}Ni_nO_{3n+1}$ samples and carried out high-pressure electrical transport measurements using DAC. The effects of external pressure and chemical doping on Ruddlesden-Popper phases are discussed, which would enrich the understanding of ongoing superconducting nickelates.

**Experiments**

Polycrystalline samples of $La_{n+1}Ni_nO_{3n+1}$ were synthesized through the solid-state reaction as described elsewhere[15, 31-33]. High-purity $La_2O_3$ (99.999%, Aladdin), and NiO (99.99%, Aladdin) are used as raw materials. We also try to prepare Sr-doping, Y-doping, and Cu-doping samples using SrO (AR, Macklin), $Y_2O_3$ (99.999%, Aladdin), and CuO (99.999%, Aladdin) and details are shown in Table S1. The powder X-ray diffraction (XRD) patterns were taken using a Bruker D2 with Cu-$K_\alpha$ radiation at room temperature. The open-source software package GSAS II is used for Rietveld refinement[34]. The chemical composition of doping samples is given by energy-dispersive X-ray spectrometry (EDS). Transport and magnetic properties under ambient pressure were measured using the Physical Property Measurement System (Dynacool, Quantum Design) and SQUID vibrating sample magnetometer (MPMS3, Quantum Design). High-pressure electrical resistance at zero magnetic field was measured in a customary cryogenic setup. Resistance measurements in the magnetic field were performed on PPMS using a nonmagnetic DAC. A cubic BN/epoxy mixture insulator layer was inserted between the BeCu gasket and electrical leads. Four Pt foils were arranged according to the van der Pauw method[35, 36]. Since we focus on polycrystalline samples, no pressure-transmitting medium is used in high-pressure transport measurements. The pressure was determined by the ruby luminescence method[37].

**Results and discussion**

Figure 1 illustrates the crystal structures of $La_{n+1}Ni_nO_{3n+1}$, which can be described as the stacking of perovskite blocks $(LaNiO_3)_n$ and rock salt (LaO) layers. With the various values of $n$ in $La_{n+1}Ni_nO_{3n+1}$, the average Ni valence state increases from +2.0 for $La_2NiO_4$ ($n =1$), +2.5 for $La_3Ni_2O_7$ ($n = 2$), to +2.67 for $La_4Ni_3O_{10}$ ($n = 3$). Among them, $La_3Ni_2O_7$ crystallizes in an orthorhombic structure with space group *Amam*. Figure 2a shows the powder XRD data and relevant Rietveld refinement of $La_3Ni_2O_7$. Except for the peaks arising from the slight NiO impurity, all peaks could be attributed to $La_3Ni_2O_7$ with the orthorhombic structure. The results of Rietveld fitting and cell parameters are summarized in Table I, which are in good agreement with those in previous reports[12, 31, 38, 39].

We have synthesized Y-doping $La_3Ni_2O_7$ (($La,Y)_3Ni_2O_7$), which induced chemical pressure by substituting the smaller Y into the La site. As shown in the inset of Figure 2b, a clear shift of the Bragg peaks to larger angles shows up, which indicates that the lattice has a little shrinkage after doping Y into the parent phase. We tried to induce more electrons/holes in the present $La_3Ni_2O_7$ phase by Cu/Sr-doping. Interestingly, chemical doping in $La_3Ni_2O_7$ not only changes the carrier density but also the crystal structure. Figures 2c and 2d show the Rietveld refinements of powder XRD of chemical doping samples and the detailed results are included in Table I. For the Sr-doping in the $La_3Ni_2O_7$, we ultimately obtained the 214 phase (($La, Sr)_2NiO_4$), although we use the same synthesis conditions. Compared with the undoped sample, the volume of $(La, Sr)_2NiO_4$ is reduced, which owns to the valence state of Ni changes from +2 ($r_{Ni}^{2+} = 0.7$ Å) to +3 ($r_{Ni}^{3+} = 0.56$ Å) to preserve electroneutrality[40]. Unlike the Sr-doping case, the final phase becomes the 4310 phase (($La_4(Ni, Cu)_3O_{10}$) after Cu-doping in the Ni site. From the EDS analysis, we indeed found the doping elements, i.e., Sr or Cu, in our synthesized samples as shown in the inset of Figures 2c and 2d. Although the key to the chemical doping-induced structural transition remains unknown, the evolution of crystal lattices together with EDS results demonstrate the successful chemical substitution in $La_3Ni_2O_7$.

Figure 3a shows the temperature dependence of the resistivity $\rho(T)$ for $La_3Ni_2O_7$ at ambient pressure, revealing a metallic ground state. The $\rho(T)$ displays an anomaly near 126 K, which indicates the formation of density wave[41]. The temperature of anomaly in our experiment is slightly higher than that in single crystal[15], likely due to the different oxygen contents between single crystals and polycrystalline samples. Further cooling the temperature, we can observe a change in the temperature coefficient d$\rho$/d$T$ from positive to negative. Figure 3b displays the temperature dependence of the magnetization of La$_3$Ni$_2$O$_7$ at 4000 Oe. The magnetic susceptibility reveals a weak temperature dependence compared with that expected for Curie-Weiss paramagnetism by moments in Ni atoms. The minimum is at around 100 K in $M(T)$, which is consistent with previous reports[15, 41]. The upturn below 40 K with cooling temperature may be related to magnetic impurities or lattice imperfections.

Next, we carried out the electrical transport experiments on La$_3$Ni$_2$O$_7$ under high pressure. Figures 3c and 3d show the temperature-dependent resistance evolution in La$_3$Ni$_2$O$_7$ up to 46.6 GPa. Application of small pressure could effectively tune the metallic ground state to the weakly insulating one, which could be ascribed to the distortion of the NiO$_6$ octahedra[42]. Upon further increasing the pressure, the resistance at room temperature begins to decrease rapidly. At around 18 GPa, a pressure-induced insulator-metal transition was obtained accompanied by a drop at low temperature. Further increasing pressure, the drop of $R(T)$ becomes more pronounced, indicating the emergence of a superconducting transition. The $T_c$, determined by the intersection of the extrapolation of normal-state resistance and superconducting transition in the resistance curve, is about 84 K at 18.1 GPa. The $T_c$ increases slightly with pressure, and a maximum reaches 86 K at $P$ = 28.9 GPa. Beyond this pressure, $T_c$ decreases and a typical domelike evolution of $T_c$ is obtained. In addition, the $R(T)$ curve reveals a typical property of a strange metal state above 150 K (Figure 3d). The experimental results have been checked and the general behaviors are similar (Figures S2 and S3) [12, 29]. It should be noted that the resistance drops by 37% at low temperature and zero resistance is not reached in the present sample, presumably associated with the variations of oxygen content in our polycrystalline samples and the pressure hydrostaticity in the DAC.

To certify whether the resistance drop is truly associated with a superconducting transition, we measured $R(T)$ at 25.1 GPa under different magnetic fields in PPMS.

Figure 4a demonstrates that the resistance drop is continuously suppressed with increasing magnetic field. The $T_c$ seems insensitive to the external magnetic field, while the superconducting transition width becomes slightly broadened, which is similar to high-$T_c$ cuprates[43, 44]. Here, $T_c$ is determined using a 90% drop in the normal-state resistance, and plots of $H_{c2}(T)$ are shown in Figure 4b. The derived upper critical field $\mu_0 H_{c2}$ as a function of temperature can be well-fitted using the empirical Ginzburg-Landau formula $\mu_0 H_{c2}(T) = \mu_0 H_{c2}(T)(1-t^2)/(1+t^2)$, where $t = T/T_c$. The extrapolated upper critical field $\mu_0 H_{c2}(0)$ of La$_3$Ni$_2$O$_7$ can reach 138 T at 25.1 GPa, which yields a Ginzburg-Landau coherence length $\xi_{GL}(0)$ of 1.57 nm. Note that $\mu_0 H_{c2}(0)$ obtained here is similar to that in La$_3$Ni$_2$O$_7$ single crystals[12, 29, 30]. This upper critical field is comparable with some high-$T_c$ cuprates, like Hg-1212(135T)[45], and YBCO (120T)[46], indicating a potential application in high magnetic fields.

Based on the above results, we can establish a $T$-$P$ phase diagram for La$_3$Ni$_2$O$_7$ as shown in Figure 5a. In the low-pressure region, the metallic behavior of La$_3$Ni$_2$O$_7$ changes to a semiconducting one. Upon further increasing the pressure, the overall magnitude of resistance is suppressed. Superconductivity was observed at around 18 GPa. The synchrotron XRD measurements demonstrate that the application of pressure induces a structural transition from the *Amam* to *Fmmm* phase, accompanied by the change of the Ni-O-Ni bond angle from 168.0° to 180° along the *c*-axis[12]. Therefore, the superconducting phase under high pressure becomes an orthorhombic structure. The superconducting transition temperature $T_c$ reaches a maximum value of 86 K within 20-30 GPa. A typical domelike evolution of $T_c$ is obtained in the pressurized La$_3$Ni$_2$O$_7$ polycrystalline sample, which is consistent with the results of single crystals[12, 29, 30]. Although the polycrystalline samples used in present experiments prevent us from observing intrinsic evolutions of anomalies in $R(T)$, it is reported that a possible density-wave-like transition[47] is suppressed by a moderate pressure before the occurrence of phase transition and superconductivity[12, 29, 30]. The rich phase diagram in La$_3$Ni$_2$O$_7$ resembles the phenomena in the cuprates [44, 48-50] and iron-based superconductors[51-56]. It remains to understand the intimate relationship between superconductivity and other ground states. This will be a stimulus for further research from an experimental and theoretical point of view.

In addition to the parent phase, we also perform measurements of the doped samples. Figure 6a shows typical $\rho(T)$ for $(La,Y)_3Ni_2O_7$, exhibiting a weakly insulating behavior[38]. Y-doping in La site in the present $La_3Ni_2O_7$ not only induces chemical pressure but also changes the ground state. We also performed magnetic susceptibility measurements for the $(La,Y)_3Ni_2O_7$ sample and the result is similar to the undoped one (Figure 6b). We measured $R(T)$ of $(La,Y)_3Ni_2O_7$ at various pressures. Figures 6c and 6d show the typical $R(T)$ curves of $(La,Y)_3Ni_2O_7$ for pressure up to 43.1GPa. Increasing pressure induces a continuous suppression of the overall magnitude of resistance. Similar to the undoped $La_3Ni_2O_7$, a resistance drop appears at about 85.6 K under 17.6 GPa, indicating a superconducting phase transition. Further increasing pressure, $T_c$ decreases slowly and the corresponding data are also summarized in the phase diagram (Figure 5b). Besides, the superconducting behavior returns to the semiconducting one when releasing pressure. Compared with the undoped sample, a larger resistance drop by 52% (Figure S4) was reached for the $(La,Y)_3Ni_2O_7$ sample, however, the zero resistance state has not been achieved yet. The Y doping content is 10% at the present stage, we expect to enhance the doping levels in further experiments.

Similar to cuprate $La_2CuO_4$, $La_2NiO_4$ is considered an antiferromagnetic (AFM) insulator[57, 58]. Since Sr-doping can induce superconductivity and reach a maximum $T_c$ of 40 K in $La_2CuO_4$[59-62], we study the transport properties of Sr-doping $(La,Sr)_2NiO_4$ at various pressures. Figure 6a shows $\rho(T)$ for $(La,Sr)_2NiO_4$ at ambient pressure. Compared with undoped $La_2NiO_4$, the magnitude of resistivity for Sr-doping $(La,Sr)_2NiO_4$ decreases significantly, and the $\rho(T)$ exhibits a weakly insulating behavior. It has been reported that the transport of the 214 phase is sensitive to oxygen content and doping composition[63-65]. AFM order in $La_2NiO_4$ could be modulated by chemical doping (Figure 6b). As shown in Figure 6c, the $R(T)$ of $(La,Sr)_2NiO_4$ exhibits a non-monotonic evolution with increasing pressure. Over the whole temperature range, the resistance decreases with pressure and reaches a minimum at about 36 GPa, then the resistance starts to increase slowly. At the present stage, the abnormal nature of transport properties in the pressurized $(La,Sr)_2NiO_4$ remains elusive. More comprehensive studies, such as the synchrotron XRD and first-principles calculations, are needed to address this issue in the future.

$La_4Ni_3O_{10}$ contains $Ni^{2+}/Ni^{3+}$ and the average valence state is +2.67, which can be regarded as 0.67 holes doped into a background of $Ni^{2+}$. Considering that such hole doping would lie far into the overdoped regime, we induced more electrons in $La_4Ni_3O_{10}$ by doping Cu on the Ni site. In contrast to undoped $La_4Ni_3O_{10}$, the $\rho(T)$ of $La_4(Ni,Cu)_3O_{10}$ at ambient pressure exhibits a metallic behavior at the high-temperature region, followed by a slight upturn at the low temperatures (Figure 7a). Figure 7b shows the magnetization of $La_4(Ni,Cu)_3O_{10}$ under magnetic field (H = 4000 Oe), a clear AFM transition is at $T_N$ = 15.5 K. The magnetization hysteresis curves measured at different temperatures are shown in the inset of Figure 7b. The field dependence of magnetization $M(H)$ at 2.5 K shows a hysteresis, while the hysteresis behavior vanishes at 200 K and the $M(H)$ curve becomes linear dependence, indicating a paramagnetic behavior. Then, we carried out transport measurements for $La_4(Ni, Cu)_3O_{10}$ under various pressures. In the low-pressure region, the $R(T)$ exhibits a weakly insulating behavior. It displays a down-turn feature at about 200 K followed by a slight upturn at the lower temperatures (Figure 7c-d). With increasing pressure, the magnitude of resistance drops dramatically, however, the $R(T)$ does not display a monotonous metallic behavior in the whole temperature range even under the maximum pressure in our research. We also studied the transport properties of the parent $La_4Ni_3O_{10}$ phase under high pressure (Figure S5), which shows similar behavior to the doping samples. In contrast to the $La_3Ni_2O_7$ family, no resistance drop or superconducting transition was observed. Further experiments up to higher pressure regions are needed to verify the possible pressure-induced superconductivity in the $La_4Ni_3O_{10}$ family.

**Conclusion**

In summary, we have performed a comprehensive high-pressure study on the electrical transport properties of the Ruddlesden-Popper Phases in DACs. Our study reveals a dome-shaped superconducting phase diagram in the pressurized $La_3Ni_2O_7$ polycrystalline samples, which is consistent with previous reports in single crystals. No signatures of superconductivity were observed in $La_2NiO_4$ and $La_4Ni_3O_{10}$ systems upon compression to 50 GPa. Further experimental studies, e.g., higher pressure measurements and more systematic chemical doping, are still needed to explore high $T_c$ superconductivity in nickelates.

**Acknowledgment**

This work was supported by the National Natural Science Foundation of China (Grant Nos. 52272265, U1932217, 11974246, 12004252), the National Key R&D Program of China (Grant No. 2018YFA0704300), and the Shanghai Science and Technology Plan (Grant No. 21DZ2260400). The authors thank the support from the Analytical Instrumentation Center (# SPST-AIC10112914), SPST, ShanghaiTech University.

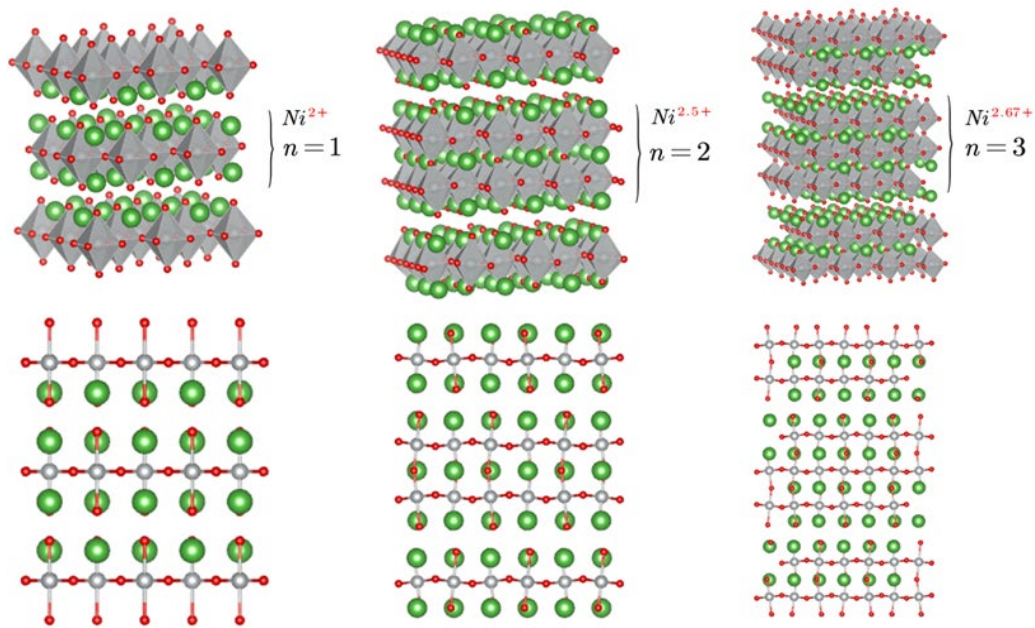

**Figure 1**. Crystal structure of Ruddlesden-Popper phases $La_{n+1}Ni_nO_{3n+1}$. Left: $n$ = 1, $La_2NiO_4$ (I4/*mmm*). Middle: $n$ = 2, $La_3Ni_2O_7$ (*Amam*). Right: $n$ = 3, $La_4Ni_3O_{10}$ ($P2_1/a$).

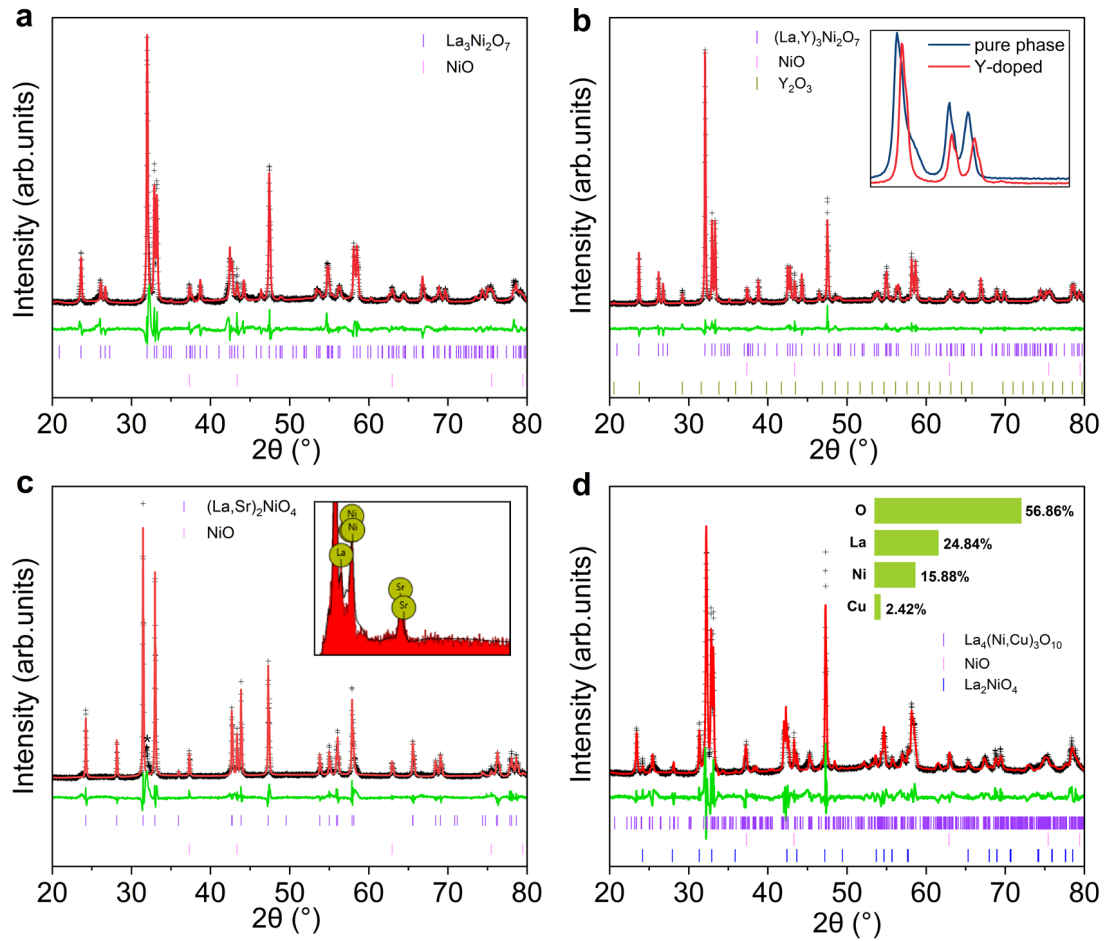

**Figure 2**: Rietveld refinement of powder X-ray diffraction data for doped Ruddlesden-Popper phases. (a) $La_3Ni_2O_7$. (b) $(La,Y)_3Ni_2O_7$. The inset shows the shift of the peaks between $(La,Y)_3Ni_2O_7$ and $La_3Ni_2O_7$. (c) $(La,Sr)_2NiO_4$. The inset shows the typical EDS. (d) $La_4(Ni,Cu)_3O_{10}$. The inset shows the element content.

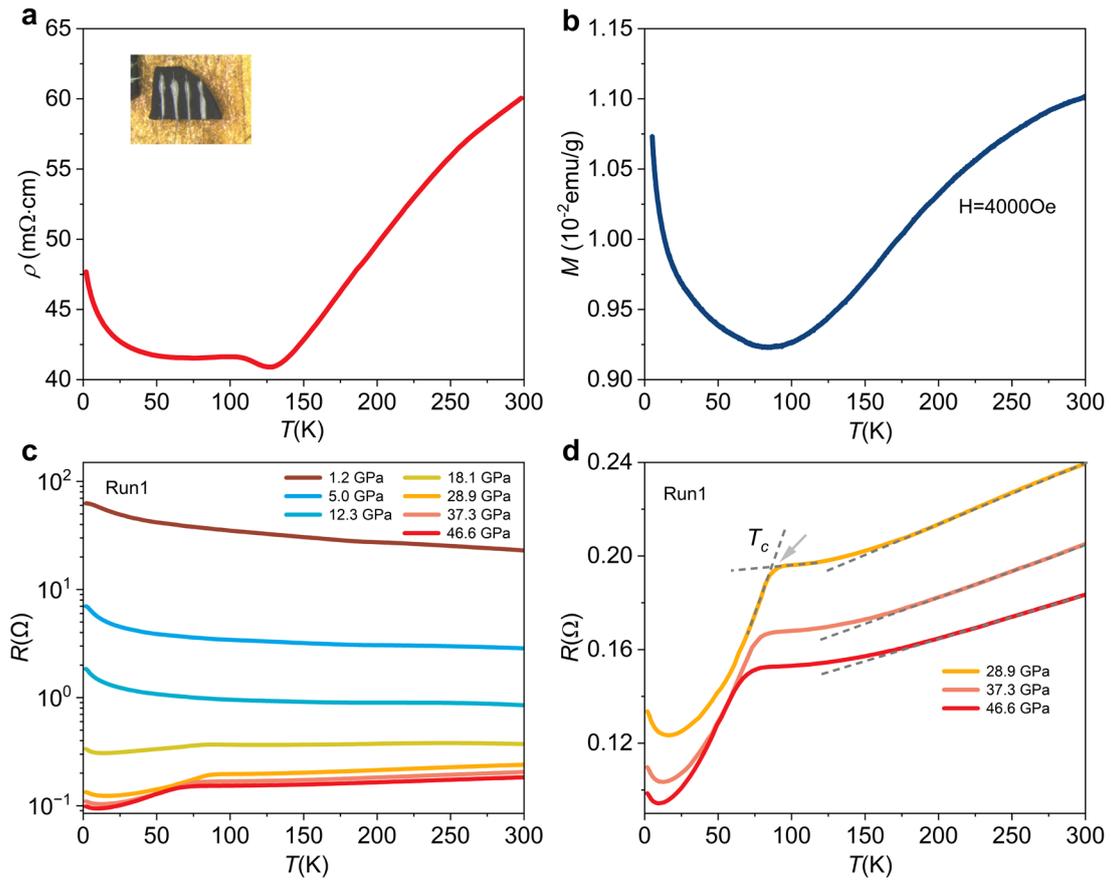

Figure 3: (a) Temperature-dependent resistivity of $La_3Ni_2O_7$ at ambient pressure. The inset shows an optical micrograph of the sample. (b) Magnetic susceptibility of $La_3Ni_2O_7$ at ambient pressure. (c)-(d) Temperature-dependent resistance of $La_3Ni_2O_7$ at various pressures. The straight line shows the linear resistance in the normal state.

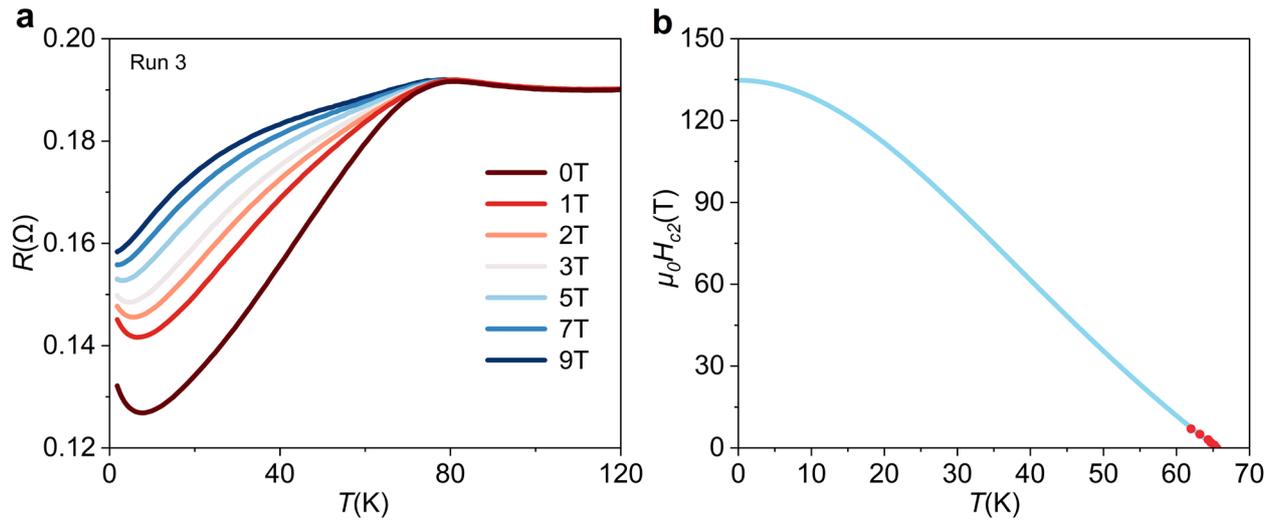

**Figure 4**: (a) Temperature-dependence of resistance under different magnetic fields for $La_3Ni_2O_7$ at 25.1GPa. (b) The upper critical field $\mu_0H_{c2}(T)$ for $La_3Ni_2O_7$ at 25.1GPa. The solid lines represent fits based on the Ginzburg–Landau (G-L) formula.

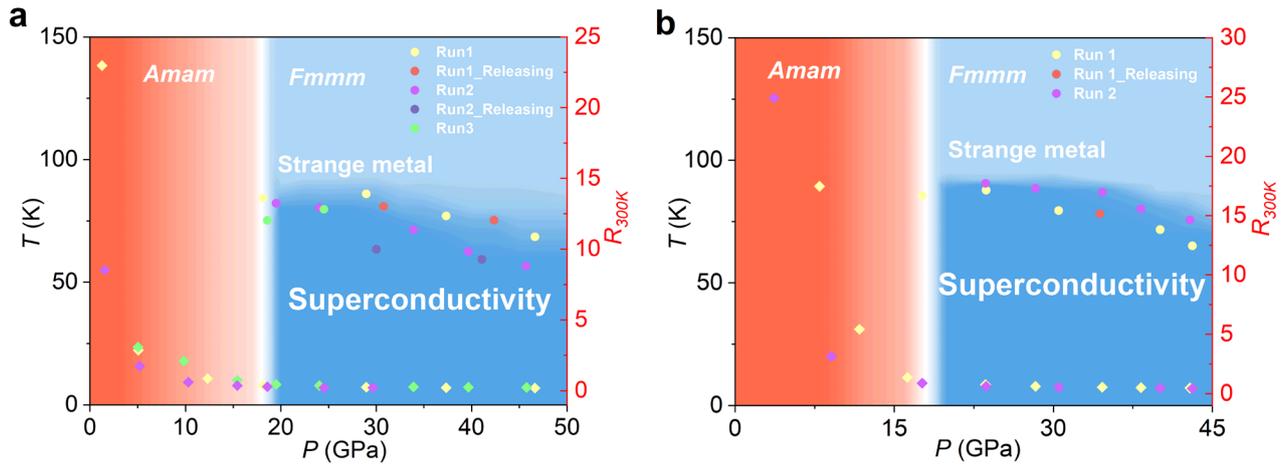

**Figure 5**: Phase diagram of La$_3$Ni$_2$O$_7$ (a) and (La,Y)$_3$Ni$_2$O$_7$ (b). The $T_c$ (circles) extracted from different runs of electrical resistivity measurements. The resistance values at 300 K (squares) are also shown. Colored areas are a guide to the eye indicating the distinct phases.

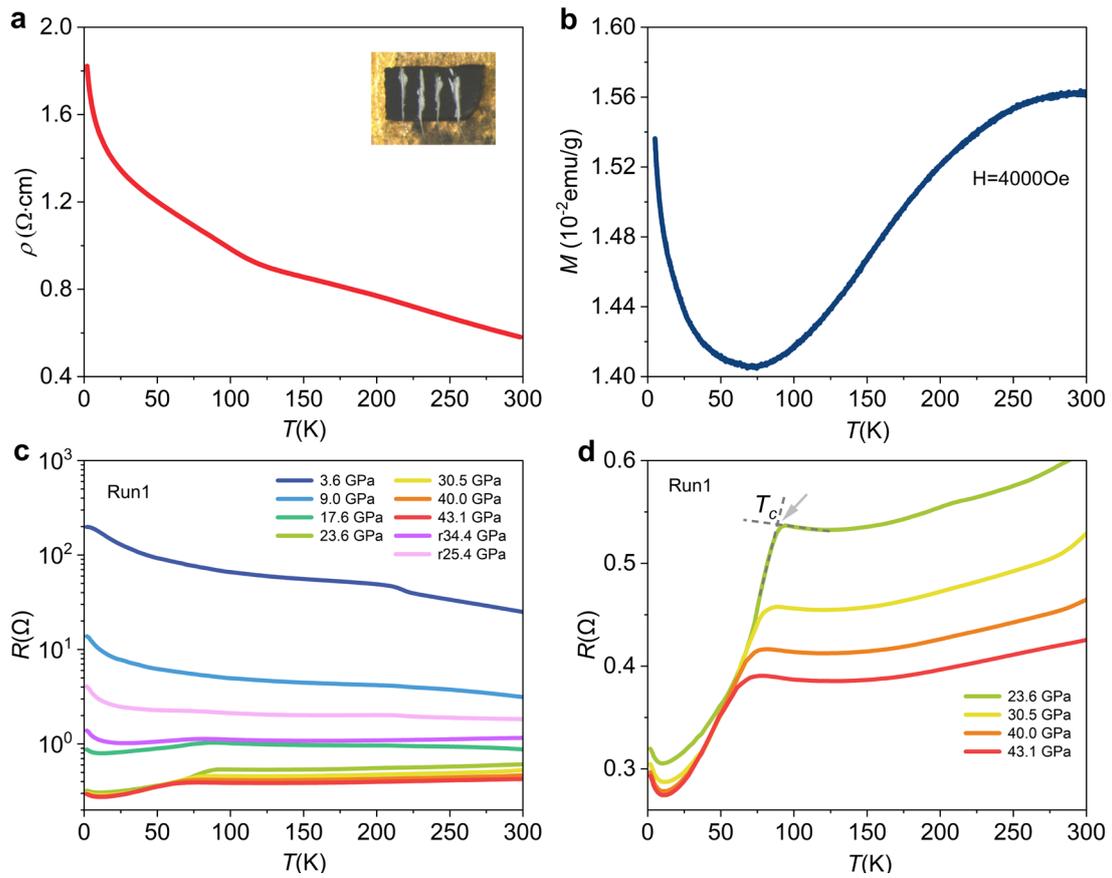

**Figure 6**: (a) Temperature-dependent resistivity of $(La,Y)_3Ni_2O_7$ at ambient pressure. The inset shows an optical micrograph of the sample. (b) Magnetic susceptibility of $(La,Y)_3Ni_2O_7$ at ambient pressure. (c)-(d) Temperature-dependent resistance of $(La,Y)_3Ni_2O_7$ at various pressures.

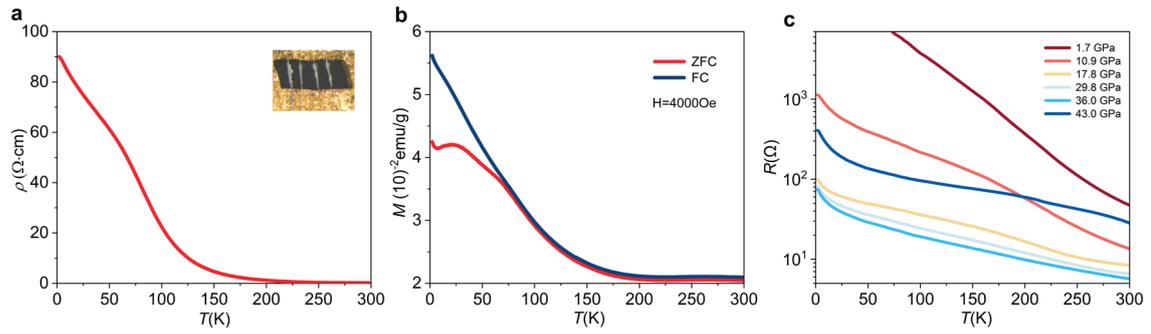

**Figure 7**: (a)Temperature-dependent resistivity of (La,Sr)$_2$NiO$_4$ at ambient pressure. The inset shows an optical micrograph of the sample. (b) Magnetic susceptibility of (La,Sr)$_2$NiO$_4$ at ambient pressure. (c)Temperature-dependent resistance of (La,Sr)$_2$NiO$_4$ at various pressures.

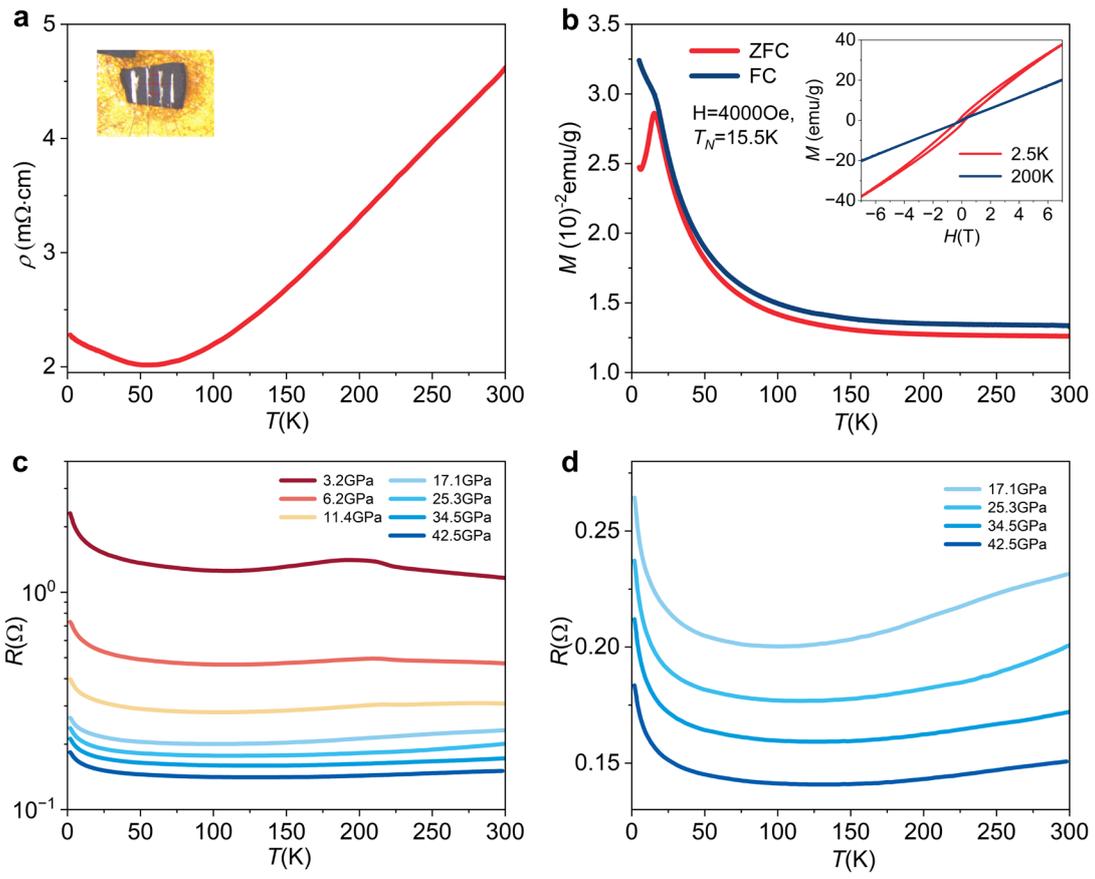

**Figure 8**: (a) Temperature-dependent resistivity of La$_4$(Cu,Ni)$_3$O$_{10}$ at ambient pressure. The inset shows an optical micrograph of the sample. (b) Magnetic susceptibility of La$_4$(Cu,Ni)$_3$O$_{10}$ at ambient pressure. The inset shows magnetization hysteresis loops of La$_4$(Cu,Ni)$_3$O$_{10}$ at 2.5 and 200 K. (c)-(d) Temperature-dependent resistance of La$_4$(Cu,Ni)$_3$O$_{10}$ at various pressures.

**Table I**: Details of the Rietveld refinement of Ruddlesden-Popper phases.

| Compound | 214 phase | | 327 phase | | 4310 phase | |
|---|---|---|---|---|---|---|
| | $La_2NiO_4$ | $(La,Sr)_2NiO_4$ | $La_3Ni_2O_7$ | $(La,Y)_3Ni_2O_7$ | $La_4Ni_3O_{10}$ | $La_4(Ni,Cu)_3O_{10}$ |
| Space group | I4/mmm | I4/mmm | Amam | Amam | $P2_1/a$ | $P2_1/a$ |
| a(Å) | 3.86459 | 3.84613 | 5.40171 | 5.39125 | 5.4198 | 5.41382 |
| b(Å) | 3.86459 | 3.84613 | 5.45365 | 5.45327 | 5.46844 | 5.45791 |
| c(Å) | 12.68898 | 12.69904 | 20.50778 | 20.48486 | 28.01741 | 28.05358 |
| α(°) | 90° | 90° | 90° | 90° | 90° | 90° |
| β(°) | 90° | 90° | 90° | 90° | 90.2410° | 90.3379° |
| γ(°) | 90° | 90° | 90° | 90° | 90° | 90° |
| V(Å³) | 189.510 | 187.854 | 604.139 | 602.254 | 830.368 | 828.916 |
| wR(%) | 6.094 | 9.613 | 8.815 | 5.427 | 6.527 | 8.421 |